 \documentclass[preprint2]{aastex}

\shorttitle {VII Zw 19}
\shortauthors{Beck et al.}

\begin{document}

\title{MERLIN and VLA Observations of VII Zw 19: Distant Cousin of M82}

 \author{Sara C. Beck\altaffilmark{1}}
\affil{Department of Physics and Astronomy\\ Tel Aviv University, Ramat Aviv,
Israel}
\email{sara@wise1.tau.ac.il}

\author{S. T. Garrington\altaffilmark{2}}
\affil{Jodrell Bank Observatory, University of Manchester\\ Macclesfield, Chesire SK11 9DL, UK}
\email{stg@jb.man.ac.uk}

\author{Jean L. Turner\altaffilmark{3}}\affil{Department of Physics and
Astronomy\\ UCLA, Los Angeles, CA 90095-1562}
 \email{turner@astro.ucla.edu}

\and

\author{Schuyler D. Van Dyk\altaffilmark{4}}
\affil{Infrared Processing and Analysis Center,  California Institute of
Technology, MS 100-22, Pasadena, California 91125}
\email{vandyk@ipac.caltech.edu}

\begin{abstract}
We have observed the starburst dwarf galaxy VII Zw 19
at subarcsecond resolution with the
MERLIN and MERLIN+EVN networks at 18 cm, with MERLIN at 6 cm, and
with the VLA in A array at 6 and 2 cm. The galaxy is detected at
all VLA wavelengths and is resolved. It is also resolved
at 18 cm by MERLIN, but is not detected by the EVN nor by MERLIN at
6 cm.

VII Zw 19 has a complex structure of nonthermal radio emission at
18 cm extended over $\sim$1200-1800 pc (4-6\arcsec ). That the EVN did not
detect this emission indicates that there are no obvious point sources that
could be radio supernovae. The 2 cm emission, 
predominantly thermal free-free emission, has a markedly
 different spatial distribution from the nonthermal emission. The
radio colors show that the galaxy contains numerous supernova
remnants, as well as many young HII regions, many of which are
optically thick at 6 cm. Two extended regions of 2 cm emission have little
18 cm flux and are probably emission from young star clusters. VII Zw 19
resembles M82 in its radio and infrared spectrum, but
the starburst region in VII Zw 19 is twice the size of the one
in M82 and twice as luminous.

\end{abstract}


\section{Introduction}
VII Zw 19 is a dwarf galaxy at a distance of 64 Mpc \citep{C91}, with the
optical-UV spectrum of a young starburst. It is classified as a blue
emission-line galaxy and a Wolf-Rayet galaxy \citep{S99}.  The optical image of the galaxy
is dominated by a strong central source with the plumes and jets suggesting
starburst winds \citep{Z}.
The Wolf-Rayet feature in the spectrum of a galaxy is a sign that many massive
stars have formed within the past 2--3 Myr.
Dwarf Wolf-Rayet galaxies are particularly interesting, because they
 have less continuous, ``background," star formation than do large spirals,
and therefore are less processed, have lower metallicity, and can be a local laboratory for studying
earlier epochs of the Universe.  When dwarf galaxies undergo a starburst, it is an
event isolated in both space (as dwarfs have no large-scale structural forces) and time, i.e.,
not confused by the ongoing star formation: The starburst emission is more ``pure" than in
spirals. The radio emission of dwarf galaxies is usually
dominated by thermal emission from the current crop of HII regions, rather than the nonthermal
emission from cosmic rays generated by supernovae from older stars,
perhaps, in part, because of the episodic nature of their
star formation activity (Deeg, Duric, \& Brinks 1996; Beck, Turner, \& Kovo 2000; however, 
magnetic fields may also play a role, cf. Turner, Ho, \& Beck 1998).

VII Zw 19 was observed at  H$\alpha$ with the Wise Observatory and
with the Very Large Array (VLA)\footnote{The VLA telescope of the National Radio
Astronomy Observatory (NRAO) is operated by Associated Universities, Inc. under a
cooperative agreement with the National Science Foundation.} at 20, 6, 3.6 and 2
cm at  $\sim$ 1\arcsec\ resolution by \citet[][]{B00}, who found that the galaxy at all
wavelengths has a strong central source in a weak envelope of
emission. VII Zw 19 has an unusually strong nonthermal radio
component for a Wolf-Rayet dwarf galaxy. The total fluxes in the
VLA maps \citep{B00} result in spectral indices of $-0.45$ between 20
and 6 cm and $-0.24$ between 6 and 2 cm, indicating that there is
both thermal (spectral index $-0.1$) and nonthermal (typical
spectral index $-0.4$ to $-1.4$) emission present. VII Zw 19 looks, in its
radio spectrum, as if it could be a transition object between the
very young starburst dwarfs, with their completely thermal or
optically thick thermal spectra, and the older starbursts, with
their plentiful supernovae. There was some indication in the
\citet{B00} images that the thermal emission came from the outer
regions of the galaxy.
However, the
arcsecond resolution of the images corresponds to $\sim$ 500 pc at 65 Mpc,
which is inadequate to resolve much structure within the starburst.

We have taken another look at the radio emission from VII Zw 19, this time at
subarcsecond resolution, with the hope of
distinguishing thermal from nonthermal emission regions, and
radio supernovae or supernova remnants (SNRs) from extended regions
of diffuse synchrotron emission.
We observed VII Zw 19 at 18 cm with the MERLIN+EVN
network in 2000 November, at 6 cm with MERLIN alone in 2001 August,
and at 20, 6 and 2 cm with the VLA in A configuration in 2002 March.
These images 
allow us to
distinguish regions that are only $\sim$50--60 pc across.

\section{Observations and Nondetections}

VII Zw 19 was observed on 2000 Nov 10 with the MERLIN 
+EVN\footnote{The European VLBI Network is a facility of European,
Chinese and other radio astronomy institutes funded by their
national research councils.} array, including the Cambridge,
Jodrell, Westerbork, Effelsburg, Mendicina and Noto antennae, and
in 2001 August with the MERLIN array alone at 6 cm (specifics are
given in Table 1). The data were reduced in the usual way with
AIPS\footnote{The Astronomical Image Processing System (AIPS) was 
developed by the NRAO.}, including phase referencing. There were two calibrators
(J0426+6825 and J0427+6821) observed, and the MERLIN data were reduced
twice, independently, such that  each reduction used only one or the other of the 
two calibrators. The differences in the fluxes from the two reductions
are less than $1\sigma$ per pixel. Images made from the two
reductions are essentially identical, which argues
strongly for the reality of the complex structures observed.

The EVN maps show a few sources in the
field, with peak fluxes up to 0.25 mJy/beam, but none of them are anywhere
near the peaks seen in the MERLIN 18 cm maps.  We conclude that they are
either background radio sources or noise. The non-detection with
EVN of a source easily detected in MERLIN argues that the emission from VII Zw
19 is so extended as to be resolved out by the EVN beam ($0{\farcs}035$=11 pc). The
rms noise in the EVN data is 0.059 mJy/beam, so, if the peak flux in the EVN beam is
reduced by a factor of 3--4 relative to the flux in the MERLIN beam, it would not be
detected.  However, the EVN beam is smaller than the MERLIN beams by factors of
20--30 in area.  Therefore, the non-detection by EVN may indicate that there are no
compact components at the 0.3 mJy/beam level, but does not rule out
extended ($\ga 0\farcs1$) sources. The MERLIN 6 cm observations
do not clearly detect the galaxy either; again, the beam at 6 cm is only one-tenth the
area of the 18 cm beam, and the flux at 6 cm is expected to be lower than at 18 cm.
This non-detection also indicates a lack of compact emission.
There are no very bright compact ($<$10--12 pc) sources,
such as radio supernovae or giant compact HII regions, detected in VII Zw 19.

The VLA observations were performed in the A configuration on 2002 March
25.  The weather was good, and 27 antennas were employed in the observations. 
For the 2 cm observations we used reference pointing and fast-switching,
with a cycle time of 2 min. Fluxes at 6 cm were based on
observations of 3C48, following the VLA 1999.2 values with adopted
fluxes of 7.91/7.98 Jy for the two IF bands. For
the 2 cm observations, which resolve out 3C48,
we adopted a flux of 0.67 Jy for the polarization calibrator 0713+438,
based on the fluxes at 6 cm, 3.6 cm, and
1 cm measured by S.~Myers \& G.~Taylor on 2002 
March 20\footnote{http://www.vla.nrao.edu/astro/calib/polar/2002/}, 
five days before
our run. We estimate that the
uncertainty in the 2 cm flux of the A array data is $\sim$5\%, based on the 
power-law fit to the spectrum; the variability of this calibrator, for which the flux
is tracked every two weeks, was less than 2\% for all of 2002, so the power-law 
fit dominates the uncertainty in the 2 cm flux scale. The uncertainty in
the flux scales based on the standard flux calibrator 3C48 is similar or less. 
Beamsizes and rms noise levels for the maps are presented in Table 1.

These A configuration data are ideal for detecting small sources,
but do not reveal extended emission well. We therefore
added to them the lower resolution VLA data from \citet{B00}. The 6
cm data of that run were obtained in B configuration, and the 2 cm
data in C configuration; those data were combined with these new A
configuration datasets to produce maps which are sensitive to
structures up to 20\arcsec\ in extent. The short 20 cm ``snapshot'' data
that we obtained for VII Zw 19 in this run are consistent with, but
noisier than, the longer integration observation in \citet{B00}.
The older data (from 1997) in \citet{B00} were also obtained in 
A configuration, so the 20 cm map derived from those data are 
shown here.

VII Zw 19 is not a very well-known radio source, and there are few
archival data that can tell us whether or not the galaxy is time-variable.
\citet {C96} measured the same 20 cm flux in 1993 with the VLA in
A configuration as did \citet{B00} in 1997.  The
only previous VLA measurement is from 1984 and was generously
given to us by G. Wynn-Williams. That program observed the galaxy
at 20, 6 and 2 cm, but used configurations with lower resolution
than did \citet{B00}. The spatial details are not comparable, but
the total fluxes are consistent with there having been no flux
variation between 1984 and 2000.

\section{Images and Radio Colors}

MERLIN detected emission from VII Zw 19 at 18 cm and the VLA
detected the galaxy at all wavelengths: 20, 6, and 2 cm. The radio images
are
shown in Figure 1. For the VLA data we show two views of the radio
emission at each wavelength. The pale grey contours represent images
with beams of 1--2\arcsec\ (FWHM) and increased sensitivity
to extended emission; the weighting (Gaussian taper of 400 k$\lambda$)
for these maps favors the
shorter VLA baselines within the combined dataset. The black
contours represent the high resolution images, which were made giving
more weight to the new A configuration data (by setting AIPS mapping 
parameter ``robust'' = 1 or 2) and which 
are sensitive to compact sources (and less so to extended emission at lower 
intensity levels). The low
resolution 20 cm image in Figure 1a is from \citet{B00}, and the
high resolution image is from the 18 cm MERLIN data. 
The MERLIN and VLA runs were, unfortunately,
on two different astrometric systems (J2000 at the VLA
and B1950 at MERLIN), so the B1950 ({\it u},{\it v}) data were converted to J2000
using the AIPS utility
UVFIX. Experience has shown that the converted data should be accurate at the
10--20 mas level, which is only one-tenth the beam size  in the highest resolution
images.
Consistent with the MERLIN and EVN images, there are no point
sources in the VLA maps, even at beam sizes of $0{\farcs}15$ (47 pc). Total
fluxes obtained using the AIPS utility IRING are listed in Table 1.
The 20 cm, 6 cm, and 2 cm fluxes from these new VLA datasets agree 
(to within 5\%) with the 1997 fluxes
from \citet{B00}, as do the fluxes from the combined 6 and 2 cm datasets in
Table 1.

\subsection {The Emission Mechanisms}

The different radio wavelengths observed are sensitive to
different emission mechanisms and physical conditions, and the
spectral index, $\alpha$, can in theory be used to separate nonthermal
sources from HII regions \citep{TH94}. The 18 and 20 cm maps are
probably dominated by nonthermal emission, the 2 cm by thermal,
and the 6 cm should be sensitive to both. Of all the observations,
the 18 cm MERLIN and 2 cm VLA have the most nearly-matched beam
sizes, $\sim$0\farcs 2--0\farcs 25. The 18 cm map, converted to J2000 and
convolved to a beam size of 0\farcs 3, is shown in
Figure 2 overlaid on a 2 cm map having the same beam.

The 2 cm and 18 cm maps are the most important here, because they
have the highest and most nearly comparable resolution and span the
longest wavelength range. In the highest resolution images
(Figures 1a and 1c) the central source and the southwest source
appear in both wavelengths. They are not identical, especially the western
filament, which splits into two in the 18 cm map. The 2 and 18 cm
maps also differ significantly southeast of the main source, where
there is a 2 cm source with no 18 cm counterpart.   The outer
parts of the galaxy appear to contain more free-free emission than
do the inner parts. These sources may be young star clusters,
whose spectra are mostly thermal, and which, in addition, may be
optically thick at 18 cm. They will be discussed further in
Section 3.3.

The differences between the 18 cm MERLIN image and the 2 cm
VLA image (Figures 1a and 1c) are also reflected in the 6 cm to 2 cm spectral
index map (Figure 1d). We constructed a spectral index map from
matched beam images, at $0\farcs3$ resolution, from the VLA
combined datasets. These maps are both in the same coordinate
system, with matched beams, have similar noise levels, and contain
short spacing information, so the spatial ({\it u},{\it v}) match, while not perfect,
is very good.
The uncertainties in $\alpha$ depend on the signal level; since
the spectral index maps are blanked when either map drops below 4
$\sigma$, we estimate that the uncertainty in $\alpha$ is
lower than 15\%. The spectral index map is shown in Figure 1d, overlaid with
the $0\farcs3$ resolution 2 cm map. The southeast and southwest
thermal filaments that are suggested by Figure 2 are evident here.
There are other possible thermal free-free sources, including the
source to the south, which have been blanked, due to weakness in
the 6 cm map.


The overall spectral index of the galaxy from the total fluxes, is $-0.42$
between 18 and 2 cm, and $-0.2$ between 6 and 2 cm. The 2 cm flux is
inconsistent with a model of a uniform nonthermal component of
spectral index $-0.7$ and an optically thin free-free component.
Either the nonthermal component has a very steep spectral index of
$-1.5$, or the 2 cm free-free emission has a significant component
with a rising flux. 

When we fold in the spatial information
contained in the 6 cm to 2 cm spectral index map in Figure 1, we see that
the nonthermal spectral index is not as steep as $-1.5$. There
are indeed spatially distinct regions of emission with a rising spectrum. 
The radio flux from this galaxy
thus appears to have a significant contribution from optically
thick thermal emission at 2 cm and from nonthermal emission at 18 cm, 
and that the
thermal emission is found preferentially to the outer edges of the 
radio source and to the south. The spatial variation of spectral index
also suggests that there may be thermal emission in the center, but
that it is confused with synchrotron emission within our 310 pc beam.


The final considerations constraining the nonthermal emission mechanism are that there is no very bright point source
and that the flux from the central region has not changed with time since 1984.  These points make it very unlikely
that VII Zw 19 contains any very luminous radio supernovae, similar to SNe 1986J or 1988Z 
(Weiler, Panagia, \& Sramek 1990; Williams et al.~2002).
The nonthermal emission must instead be produced by remnants of supernovae.

\subsection{An Extended and Mature Starburst Region}

The compact nonthermal emission, as traced by the 18 cm MERLIN map, which had $\theta_{max}=6.5\arcsec$, is
confined to less than 1\arcsec\  (310 pc) and has a total mapped flux
of 7.5--12.5 mJy, depending on which calibrator was used (see
above). The total mapped flux in the 20 cm VLA maps ($\theta_{max}=38\arcsec$) is 
24 mJy, which
corresponds to 22 mJy at 18 cm for a spectral index of $-0.7$. So
about half to two-thirds of the total 20 cm flux from VII Zw 19 is extended
emission from outside the central source.  The high spatial
resolution MERLIN observations  have resolved out
the most extended component of the long-wavelength emission.


The radio emission in VII Zw 19 differs from the young
star formation sources in the dwarf galaxies NGC 5253 \citep{T98}
or II Zw 40 \citep{B02}, which have almost no nonthermal emission.  
The 18 cm flux of the central region corresponds to the equivalent of 2500 Cas A SNRs
\citep{B77}, comparable to what is deduced for the starburst in M82. The starburst in VII Zw 19 has evolved into the stage of plentiful supernovae and/or the means of expressing this
state in the form of synchrotron emission.
The thermal emission at 2 cm is, adjusted for distance, about 100 times as strong as that from the purely thermal, optically thick
HII region excited by the super star cluster in NGC 5253 \citep{T98}.  Therefore, what roughly 
emerges is a picture of the central 310 pc of VII Zw 19 containing
a few thousand SNRs and a few hundred dense young HII regions.

\subsection{The Extra-Nuclear Thermal Sources: Bright Star Clusters?}

These high resolution maps of VII Zw 19 paint a remarkable picture of a galaxy where star formation is simultaneously
extended and compact. The total extent of the starburst is 1.2 kpc, yet there are distinct regions of 
nonthermal and thermal
emission, of which the nonthermal are the older and more evolved and the purely thermal the youngest.  We now consider in detail
the thermal sources, the youngest and, thus, the current starburst regions.  We use 
the predominance of emission at 2 cm as the
signpost for thermal sources, although this may be an oversimplification; the central core of the galaxy has a nonthermal spectral
index, yet it probably contains both thermal and nonthermal emission.

We therefore isolate the current starburst regions from the
high resolution 2 cm map (Figure 1c). In the southeast
there is a clear peak at RA(J2000)=04$^h$ 40$^m$ $47{\fs}369$, 
Dec(J2000)=+67$\arcdeg$ 44$\arcmin$ $09{\farcs}13$, with a peak flux
of 0.3 mJy/beam, and total flux of 0.4 mJy, which corresponds to 
${\rm N_{Lyc}} \sim 1.5\times 10^{53}$ s$^{-1}$ in a region roughly 110pc $\times$ 40 pc
in size (the peak is clearly extended; the source is assumed to be Gaussian and 
deconvolved from the beam). A weaker peak is found at 04$^h$ 40$^m$ $47{\fs}45$, 
$+67\arcdeg$ $44\arcmin$ $09{\farcs}10$,
with peak 0.24 mJy/beam, and total flux 0.3 mJy, which corresponds
to ${\rm N_{Lyc}} \sim 1.0\times 10^{53}$ s$^{-1}$ in a region roughly 75 pc $\times$ 30 pc.
The total flux at 2 cm in the southeast portion of the source (which appears as
a single source in the lower resolution image of Figure 1d) is 0.9 mJy, or
${\rm N_{Lyc}} \sim 3.3\times 10^{53}$ s$^{-1}$ in a region about 150 pc ($\sim 0\farcs5$)
across.

The filament in the southwest is extended and thermal. The total flux in this
filament is about 0.5 mJy, corresponding to ${\rm N_{Lyc}} \sim 2\times 10^{53}$ s$^{-1}$, 
covering a total extent of $\sim$350 pc with no clear structures down to
our smallest 2 cm beam of $0{\farcs}2$ ($\sim$60 pc).

There are also regions of thermal emission, indicating starburst activity, quite distant from
the center. Roughly 600 pc to the south of the
main radio source is an extended region containing 0.3 mJy of thermal
flux, for a total ${\rm N_{Lyc}} \sim 1.1\times 10^{53}$ s$^{-1}$, or the equivalent of
$1.1\times 10^4$ O7 stars. A compact source is visible about 0.5 kpc
to the northeast of the central source, at
04$^h$ 40$^m$ $47{\fs}59$, $+67\arcdeg$ 44$\arcmin$ $10{\farcs}4$. 
This compact source has a peak of 0.24 mJy/beam. It is unresolved, and 
therefore less than 45 $\times$ 34 pc in size, with 
${\rm N_{Lyc}} \sim 8.5\times 10^{52}$ s$^{-1}$.
This source by itself is an impressive center of star formation, containing the equivalent of
8500 O7 stars!

The thermal radio emission we estimate for all the sources we have identified as being true starburst
corresponds to total ionization of ${\rm N_{Lyc}} \sim 7.3\times 10^{53}$ s$^{-1}$, or the equivalent of
$7.3\times10^4$ O7 stars.  Using the relation 
${\rm L_{OB}}=2$--$3\times 10^{-44}\ {\rm N_{Lyc}}$, we find the expected
luminosity of the exciting OB stars to be 1.4--$2.2 \times 10^{10}\ L_\odot$. The total infrared luminosity of
this galaxy from the IRAS fluxes is $7.3\times10^{10}\ L_\odot$, so we see that a substantial fraction of
the total luminosity comes from the compact starburst sources listed here.


\section{Conclusions: Twice the M82 Starburst at 64 Mpc}

We have emphasized what is unusual about VII Zw 19:  That it is a
dwarf galaxy with a nonthermal spectrum, that it contains an extended and highly
structured region of  nonthermal emission, and that it also holds regions of optically thick thermal
emission spatially distinct from the nonthermal sources.   Although unusual, VII Zw 19 is not unique. In
many respects it is remarkably similar to the well-known starburst
galaxy, M82 \citep{G96}.  If M82 were at 64 Mpc, instead of 3.3 Mpc, it would
have fluxes of 22 mJy, 4.5 mJy, and 0.15 Jy at 20 cm, 2 cm and
12$\mu$m, respectively, quite similar to the 24 mJy, 9 mJy and 0.29 Jy of VII Zw 19, and
it would have about half the linear extent in the radio ---700 pc as opposed to 1200 pc.
VII Zw 19 is bluer in the infrared than M82, and it has about twice the thermal
flux (i.e., twice the ${\rm N_{Lyc}}$ or number of equivalent O stars).
The main observational difference between the two galaxies are
that the starburst region in M82 is more dominated by
synchrotron emission.  M82 does not have
the equivalent of the two extended regions of optically thick thermal emission
regions seen at 2 cm in VII Zw 19.  
If M82 were at the distance of VII Zw 19, then its thermal sources could not be
separated from the nonthermal at the resolution of the images presented here.  
The two giant thermal
clusters are probably responsible for the excess at 2 cm and for
the infrared flux in VII Zw 19, as compared to M82. 
The two giant
thermal clusters may indicate that the star formation process in VII Zw 19 
is not coeval (i.e., with
the entire central region the same age), but that star
formation began at an earlier epoch in the main complex
and only more recently in the
thermal clusters. However, there may be other reasons that VII Zw 19, overall,
but particularly in the outer ``thermal" regions,  is
deficient in synchrotron emission as compared to M82.

One possible difference between M82 and VII Zw 19 is the environment.
M82 is part of a group of galaxies, interaction with
which has apparently triggered the starburst activity \citep{Y94}, while VII
Zw 19 is isolated.  What, then, triggered the starburst in VII Zw 19? A
possible explanation is that VII Zw 19 accreted another dwarf
galaxy which is no longer visible as a separate object. This is
the process believed to have started the activity in He 2-10, for
example \citep{K95}. If VII Zw 19 underwent such an accretion event in the
last 100 million years, it would have disturbed the kinematics of
the molecular and atomic gas in a fashion that should still be
observable. 

VII Zw 19 is a challenge to map and study properly. Its distance of
64 Mpc reduces the effective spatial resolution by a factor of 20 and the measured
fluxes by a factor of 400, compared to M82. The active region covers 1.2 kpc, but it contains
regions of distinct properties and histories on size scales down to our resolution limit
of 45 pc. It is a tribute to current sub-arcsecond imaging
methods that we have been able to study the structure at all. 
This galaxy is much more complicated than most starburst dwarfs, and its evolution
and star formation history remain mysterious.


\acknowledgments

Work at Jodrell Bank and the EVN was supported by the EC Access to
Research Infrastructures Programme, Contract
No.HPRI-CT-2001-00142. SCB thanks Mike Garret and Denise Gabudza
for their patient help and the staff of Jodrell Bank and JIVE for
support. JLT acknowledges the support of NSF Grant 0307950. This
work has made use of the NASA/IPAC Extragalactic Database (NED)
which is operated by the Jet Propulsion Laboratory, California
Institute of Technology, under contract with the National
Aeronautics and Space Administration, and of NASA's Astrophysics
Data System.

\clearpage
\begin{deluxetable}{lcccccc}
\tabletypesize{\scriptsize}
\tablenum{1}
\tablewidth{0pt}
\tablecaption{VII Zw 19 Radio Data}
\tablehead{
\colhead{$\lambda$}
&\colhead{Telescope}
&\colhead{rms}
&\colhead{Beam, p.a.}
&\colhead{$\theta_{max}^a$}
&\colhead{Peak Flux}
&\colhead{Total Mapped}
\\
\colhead{}
&\colhead{}
&\colhead{}
&\colhead{}
&\colhead{}
&\colhead{}
&\colhead{Flux$\rm ^{a,b}$}
\\
\colhead{}
&\colhead{}
&\colhead{(mJy/bm)}
&\colhead{(\arcsec), (\arcdeg)}
&\colhead{(\arcsec\ FWHM)}
&\colhead{(mJy/beam)}
&\colhead{(mJy)}
\\
}
\startdata
18~cm&MERLIN&0.11&0.30, \nodata & 6.5 &1.06&10$\pm$4\\
18~cm&MERLIN&0.10&0.18 $\times$ 0.18, 0$\rm^c$&6.5 & 0.9
&14$\pm$4\\
6~cm&MERLIN&0.10& 0.046 $\times$ 0.14, $-35$ &1.2 & 0.4
&\nodata \\
20~cm&VLA&0.08& 1.17 $\times$ 0.97, 14 & 35 & 5.4 & 24$\pm$2\\
6~cm& VLA& 0.04 & 0.45 $\times$ 0.34, 15 & 10 & 0.91 & 11$\pm$1\\
2~cm& VLA & 0.03 & 0.27 $\times$ 0.22, 25 & 4 & 0.35 &  8.8$\pm$2\\

\enddata

\tablenotetext{a}{$\theta_{max}$ is the maximum sizescale that is
well sampled by these images. Fluxes and peak fluxes are therefore
lower limits to the total flux.}

\tablenotetext{b}{Total fluxes are obtained by integrating over
a circular aperture centered on the peak. The large uncertainties at 18 cm are due to the 
use of two different flux calibrators, as discussed in the text.}

\tablenotetext{c} {The two 18 cm maps were made with different constraints in the mapping: The larger beam is naturally weighted, and the smaller beam was imposed in the AIPS utility IMAGR to be a symmetric beam close to the $0{\farcs}17 \times 0{\farcs}126$ beam of a ``robust''=1 map.}

\end{deluxetable}

\clearpage

\clearpage

{Figure 1. Maps of VII Zw 19, with beam sizes and wavelengths on each map. Contour levels 
are $\pm 2^{n/2}$, where $n$=0, 1, 2..., times the given contour unit, which is 0.12 mJy/beam,
unless otherwise specified. Contours begin at $\sim 4$--$6\sigma$.
(a) 18 cm MERLIN map of VII Zw 19 ({\it black contours}) atop the 20 cm VLA map 
({\it light grey contours}).  Beam sizes are $0{\farcs}18 \times 0{\farcs}18$ for the 
MERLIN map, and $1{\farcs}2 \times 1{\farcs}0$ (position angle p.a.=14\arcdeg)
for the VLA map. The MERLIN map has a contour unit of 0.3 mJy/beam (2.5 $\sigma$).
Peak flux densities are 5.4 mJy/beam for the VLA image and 0.91 mJy/beam for the 
MERLIN image.
(b) 6 cm VLA cm maps: Low resolution image ({\it grey}), with beam of 
$1{\farcs}3 \times 1{\farcs}1$ (p.a.= $-11\arcdeg$); high resolution image ({\it black}), 
with beam of $0{\farcs}45 \times 0{\farcs}34$ (p.a.= 35\arcdeg). Peak
flux densities are 3.7 mJy/beam for the low resolution image and 0.91 mJy for the
high resolution image.
(c) 2 cm VLA maps: Low resolution image ({\it grey}), with a beam of $1{\farcs}2 \times 
1{\farcs}0$ (p.a.=52\arcdeg); high resolution ({\it black}), with a beam of 
$0{\farcs}27 \times 0{\farcs}22$ (p.a.=25\arcdeg). Peak flux 
densities are 1.5 mJy/beam for the low resolution image
and 0.35 mJy/beam for the high resolution image.
(d) Spectral index map in color, with 2 cm map in {\it solid contours}. Spectral index map is 
color contoured in multiples of $\pm$0.2; the zero contour is the line between {\it yellow\/}
and {\it green}, with positive spectral indices toward the {\it red}. The 2 cm map is used
in the spectral index computation, convolved to a beam of $0{\farcs}45$  to match 
the 6 cm map. Peak flux is 0.57 mJy/beam for the 2 cm map
(the corresponding peak, not shown, in the 6 cm map is 0.70 mJy/beam). 
}

{Figure 2. The 18 cm MERLIN data, converted to J2000 and mapped to match the beam of the
2 cm VLA map, is shown in {\it blue contours\/} and the 2 cm VLA map, 
is overlaid in {\it red contours}.  The beam sizes are $0{\farcs}3$; 
the contour units are 0.12 mJy/beam for the 2 cm
map and 0.24 mJy/beam for the 18 cm map, with levels $\pm 2^{n/2}$, where 
$n$=0, 1, 2..., times the given contour unit. The VLA map
contains short spacing information and is sensitive to structures up to $\sim$20\arcsec;
the MERLIN map is sensitive only to structures $<4$\arcsec\ in extent.



\begin{thebibliography}{}
\bibitem[Baars et al. (1977)]{B77} Baars, J.W.M., Genzel, R., Pauliny-Toth, I.I.K. \& Witzel, A.
1977, \aap,61,99
\bibitem[Beck et al. (2000)]{B00} Beck, S. C., Turner, J. L., \& Kovo, O. 2000, \aj, 120, 244
\bibitem[Beck et al. (2001)]{B01} Beck, S.C., Turner, J.L., \& Gorjian, V. 2001, \aj, 122, 1365
\bibitem[Beck et al. (2002)]{B02} Beck, S.C., Turner, J. L., Langland-Shula, L. E., Meier, D. S., Crosthwaite, L.P. \& Gorjian, V. 2002, \aj, 124, 2516
\bibitem[Conti (1991)]{C91} Conti, P.S. 1991, \apj, 377, 115
\bibitem[Condon et al. (1996)]{C96} Condon, J. J., Helou, G., Sanders, D. B., \& Soifer, B. 1996,
\apjsupp, 103, 81
\bibitem[Deeg et al. (1996)]{D96} Deeg, H.-J., Duric, N., \& Brinks,  E. 1996, \aap, 323, 323
\bibitem[Golla et al. (1996)]{G96} Golla, G., Allen, M.L., \&Kronberg, P. 1996, \apj, 473, 244 L137
\bibitem[Kobulnicky et al. (1995)]{K95} Kobulnicky, H.A., Dickey, J.M., Sargent, A.I., Hogg, D.E., \& Conti, P.S. 1995, \aj, 110, 116
\bibitem[Schaerer et al. (1999)]{S99} Schaerer, D., Contini, T., \& Pindao, M. 1999, \aaps, 136, 35S
\bibitem[Turner \&  Ho (1994)]{TH94} Turner, J. L., \& Ho, P. T. P. 1994, \apj, 421, 122
\bibitem[Turner et al. (1998)] {T98} Turner, J.L., Beck, S.C., \& Ho, P.T.P. 1998, \aj, 116, 121
\bibitem[Weiler, Panagia, \& Sramek (1990)] {WPS90} Weiler, K.W., Panagia, N, \& Sramek, R.A. 1990, \apj, 364, 611
\bibitem[Williams et al. (2002)]{W02} Williams, C.L., Panagia, N., Van Dyk, S.D., Lacey, C.K., Weiler, K.W., \& Sramek, R.A. 2002, \apj, 581, 396
\bibitem[Yun et al. (1994)]{Y94} Yun, M.S., Ho, P.T.P., \& Lo, K.Y. 1994, \nat, 372, 530
\bibitem[Zwicky (1967)]{Z} Zwicky, F. 1967, \aj, 372, 530
\end{thebibliography}
\end{document}